\documentclass[prl,aps,showpacs,twocolumn,groupedaddress,superscriptaddress]{revtex4}
\usepackage{amsmath}
\usepackage{graphicx}
\usepackage{mathrsfs}
\usepackage{textcmds}
\allowdisplaybreaks[1]
\newcommand{\be}{\begin{equation}}
\newcommand{\ee}{\end{equation}}
\newcommand{\bea}{\begin{eqnarray}}
\newcommand{\eea}{\end{eqnarray}}
\newcommand{\ket}[1]{\left|#1\right\rangle}
\newcommand{\bra}[1]{\left\langle #1\right|}

\newcommand{\expect}[3]{\left\langle #1 | #2 | #3 \right\rangle}

\newcommand{\bc}{\begin{center}}
\newcommand{\ec}{\end{center}}

\renewcommand{\(}{\left(}
\renewcommand{\)}{\right)}
\renewcommand{\[}{\left[}
\renewcommand{\]}{\right]}
\newcommand{\forget}[1]{}

\newcommand{\re}{{\rm e}}
\newcommand{\ri}{{\rm i}}
\begin{document}
\title{Generation of Werner states via collective decay of coherently driven atoms\footnote{Part of this work was presented at International Conference on Quantum Optics, Minsk, Belarus, May 30-June 2 (2004) and International Conference on Uncertainty Relations and Squeezed States, Besancon, France, May 2-6 (2005).}}
\author{Girish S. Agarwal\footnote{On leave of absence from Physical Research Laboratory, Navrangpura, Ahmedabad, 380 009, India.}}
\email{agirish@okstate.edu}
\affiliation{Dept.~of Physics, Oklahoma State University, Stillwater, Oklahoma 74078-3072} 
\author{Kishore T.  Kapale}
\email{Kishor.T.Kapale@jpl.nasa.gov}
\affiliation{Jet Propulsion Laboratory, California Institute of Technology,
Mail Stop 126-347, 4800 Oak Grove Drive, Pasadena, California 91109-8099}
\date{\today}

\begin{abstract}
We show deterministic generation of Werner states as a steady state of the  collective decay dynamics of a pair of neutral atom coupled to a leaky cavity and strong coherent drive.  We also show how the scheme can be extended to generate $2N$-particle analogue of the bipartite Werner states.
\end{abstract}
\maketitle

Entanglement, one of  the most striking features of quantum physics, is a great resource  for the modern fields of quantum information and quantum computing~\cite{Nielsen:2000}. Entanglement makes possible novel quantum communication protocols such as quantum teleportation~\cite{Bennett:1993} and quantum dense coding~\cite{Bennett:1992} and is at the heart of quantum cryptography~\cite{Ekert:1991}.

Among several bipartite entangled states, Werner states~\cite{Werner:1989} provide the simplest example of  mixed states possessing entanglement. Werner states, in fact,  signify a wide class of entangled states with a varying degree of correlations depending on a single parameter.  Assuming a general mixed state density matrix $\rho_{\rm M}$ for a bipartite two-state system, the entanglement fidelity can be expressed through $F=\bra{\Psi_{\rm s}}\rho_{\rm M}\ket{\Psi_{\rm s}}$, where $\ket{\Psi_{\rm s}}$ is a singlet state---of the Bell type---of the bipartite system.  Depending on this fidelity, the Werner state demonstrates classical or quantum nature. It has been shown by Bennett {\it et al.}~\cite{Bennett:1996} that for $F>1/2$ the mixed state can be purified to obtain a pure singlet state, thus can be taken to possess non-local character~\cite{Popescu:1994}. However, for $F<1/2$ it is completely classical and can be generated by using an initially unentangled particles. Nevertheless, it has been realized~\cite{Bennett:1996} that only when $F>(2 + 3 \sqrt{2})/8\approx 0.78$ the Werner states violate the Clauser-Horne-Shimony-Holt inequality~\cite{Clauser:1969} and demonstrate quantum correlations. 
Experimentally, Werner states have been generated so far with only photonic qubits~\cite{exp}. Here, for the first time we propose a scheme for generation of Werner states with neutral atoms. The proposal can be, fairly easily, extended to trapped ions and it is well-within the realm of current  experiments. The scheme is based on collective decay dynamics of two atoms coupled coherently to a strong drive field. As the Werner states are obtained as steady state solutions of the decoherent dynamics, they are naturally stable and long-lived. The model can be extended to a $2N$-particle system to obtain a multiparticle analogue of the usual bipartite Werner state as we discuss in detail.

To throw some light on the collective dynamics of a system of two-level atoms we first consider a simple case of two atoms interacting with a leaky cavity with the decay rate $\kappa$. The atom-cavity interaction can be described through the Hamiltonian
\begin{equation}
\mathscr{H} =   \hbar \mathscr{G} \( S_1^{-} \cos{\xi} + S_2^{-} \sin{\xi} \)a^{\dagger} + \text{H. c.}\,,
\end{equation}
where $\mathscr{G}$ is the atom-cavity coupling Rabi frequency.

The cavity field has the frequency $\omega_0$ and $a$ is its annihilation operator. $S^{-}_i=\ket{g_i}\bra{e_i}$ are the atomic operators. The terms $\cos{\xi}$ and $\sin{\xi}$ are introduced to incorporate the modal dependence of the cavity field and the relative position of the two atoms. 
The density matrix dynamics of the complete atom-cavity system is given by
\begin{align}
\dot{\rho}_{\rm ac} =-\frac{\ri}{\hbar} [\mathscr{H} , \rho] - \kappa \(a^\dagger a \rho_{\rm ac}- 2 a\rho_{\rm ac}a^\dagger + \rho_{\rm ac} a^\dagger a \)\,.
\label{Eq:rhodot2}
\end{align}
Using the rotating wave approximation, and after tracing out the field degrees of freedom the density matrix equations for the atoms can be written as
\begin{equation}
\dot{\rho} = - \Gamma \( R^{+} R^{-} \rho - 2 R^{-}\rho R^{+} + \rho R^{+}R^{-}\)\,,
\label{Eq:dotrho}
\end{equation}
where $R^- = S_1^{-} \cos{\xi}+ S_2^{-} \sin{\xi}$ and $\Gamma = {\mathscr{G}^2}/{\kappa}$. We have assumed that $\mathscr{G} \ll \kappa$ (bad-cavity limit) while arriving at the above equation. The density matrix equation~\eqref{Eq:dotrho} closely resembles the one describing spontaneous decay of an atom with the rate $2 \Gamma$, except that that atomic operators $R^\pm$ are collective and thus it describes collective decay dynamics of a system of atoms. 
The steady state solution of the above matrix equation can be readily verified to be
\begin{equation}
\rho=D (R^{-})^{-1}(R^{+})^{-1}\,,
\label{Eq:Mixed}
\end{equation}
provided that the determinant of the operator matrix $R^{-}$ is non-zero and its inverse exists, with $D$ being the appropriate normalization. In case the determinant is zero, corresponding to the existence of zero eigenvalues $R^{-}\ket{\Psi_0} = 0$, the solution of Eq.~\eqref{Eq:rhodot2} is given by
\begin{equation}
\rho=\ket{\Psi_0}\bra{\Psi_0}\,.
\end{equation}
These general considerations hold, immaterial of the actual form of the operators $R^{-}$ and $R^{+}$, so long as they satisfy Lie algebra similar to the spin operators. 

It can be readily verified that the zero eigenenergy states of $R^{-}$ are $\ket{\psi_g}= \ket{g_1, g_2}$ and
$\ket{\psi_{\text{E}}}=\(\cos{\xi}\ket{g_1, e_2}-\sin{\xi} \ket{e_1, g_2}\)$.
The state $\ket{\psi_g}$ is a separable state and $\ket{\psi_{\text{E}}}$ is similar in form to one of the maximally entangled Bell-states.
It can be noted, however, that starting with the initial state $\ket{\psi_{g}}$,  the entangled state $\ket{\psi_{\text{E}}}$ could not be reached in the steady state. Thus, the initial state needs to be asymmetric i.e., with one of the atom in the excited state. For example, with the initial state $\ket{e_1, g_2}$, which can  be written as
\begin{align}
\ket{g_1,e_2} &\equiv \alpha [\cos{\xi} \ket{g_1, e_2} - \sin{\xi}\ket{e_1, g_2}] 
\nonumber \\
&\quad+ \beta [\sin{\xi} \ket{g_1, e_2} + \cos{\xi}\ket{e_2, g_2}]\,,
\end{align}
with $\alpha = \cos{\xi}$  and $\beta = \sin{\xi}$; the steady state density matrix is given by
\begin{align}
\rho = |\alpha|^2\ket{\psi_{\text{E}}}\bra{\psi_{\text{E}}}
+ |\beta|^2\ket{g_1, g_2}\bra{g_1, g_2}.
\label{Eq:mixed2}
\end{align}
Thus, the entangled state $\ket{\psi_{\text{E}}}$ is generated with the probability \mbox{$|\alpha|^2=\cos^2{\xi}$.}  Nevertheless maximal incoherent mixing is observed  for \mbox{${\xi} = \pi/4, 3 \pi/4, \dots$}.

Our calculations further suggest that even good cavities can be used to generate mixed-state entanglement at steady state. Here we use the master equation~\eqref{Eq:mixed2} and solve it for the time evolution of the density matrix of the combined system of the two atoms and the cavity mode. To illustrate, starting with an initial state $\ket{e_1,g_2}$, with no photons in the cavity mode, mixed-entanglement of the form~\eqref{Eq:mixed2} can be generated at the steady state provided $\kappa\neq0$. Thus, the ideas presented so far can also be applied to two atoms trapped in a good-cavity.

Thus, we have seen how collective decay dynamics of a system of two atoms can give rise to entanglement. In the following discussion we add a resonant coherent drive field in order to obtain extra handle over the steady state density matrix. We assume a general collective-decay mechanism which we have shown how to to obtain in a leaky cavity for example.

The dynamics of a system of  two level atoms driven coherently by a classical field (Rabi frequency $|\Omega|\re^{-\ri \phi}$) on-resonance and experiencing a collective decay at the rate $2\Gamma$ is governed by\cite{Agarwal:1990}
\begin{align}
\dot{\rho} &=  \ri\,  |\Omega|\, [ \re^{\ri \phi} S^{+} +  \re^{-\ri \phi} S^{-}, \rho] \nonumber \\
&- \Gamma(S^{+}S^{-}\rho - 2 S^{-}\rho S^{+}+ \rho S^{+}S^{-})\,.
\label{Eq:dotrhoN}
\end{align}
Here, $S^{\pm}$ are the collective atomic operators; they can be thought of as the operators for a  spin $N$ particle for a total of $2N$ particles in the system. These collective operators are very similar to the regular spin operators except for the absence of $\hbar$ dependence of their eigenvalues unlike the actual spin systems.  

Introducing a new operator defined as \mbox{$R^{-} = S^{-} + \ri\, (|\Omega|/\Gamma) \re^{\ri \phi}$}~\cite{Puri:1979,Narducci:1978},  the density matrix equation~\eqref{Eq:dotrhoN} takes the same form as Eq.~\eqref{Eq:dotrho}. Therefore, the steady state solution, which is analogous to Eq.\eqref{Eq:Mixed},  can be readily arrived at.  In the strong-field limit with an initially symmetric state of the atomic spins the steady state solution is given by
\begin{equation}
\rho(\infty)=\frac{1}{2 S + 1} \sum_{m=-S}^{S} \ket{S, m}\bra{S,m}\,,
\label{Eq:steadystate}
\end{equation}
which is a mixed state with all possible $S^{z}$ eigenstates equally occupied.  Moreover, the state $\ket{S=0,m=0}$ is the zero eigenvalue state giving a pure state solution if the initial state overlaps with it.

In the following, we discuss how the collective decay dynamics can be exploited to generate a bipartite Werner state of a pre-chosen fidelity.
We start with a general initial state $\sin\theta \ket{e_1, g_2} +\cos\theta \ket{g_1, e_2}$, for a bipartite system.  It can be readily rewritten as
\begin{equation}
\ket{\psi_0}=\[\( \frac{\sin \theta + \cos \theta}{\sqrt{2}} \) \ket{\Psi^{+}}+ \( \frac{\sin\theta - \cos\theta}{\sqrt{2}}\)\ket{\Psi^{-}} \] 
\label{Eq:initial}
\end{equation}
in terms of two of the Bell States
\begin{align}
\ket{\Phi^{\pm}}= \frac{\ket{e_1 e_2} \pm \ket{g_1 g_2}}{ \sqrt{2}},\,\,\,
\ket{\Psi^{\pm}}= \frac{\ket{e_1 g_2} \pm \ket{g_1  e_2}}{ \sqrt{2}}\,.
\label{Eq:BellStates}
\end{align} 
The Bell states can be readily written in terms of the collective spin states---$\ket{S,m}$---as
\begin{align}
\ket{\Psi^{+}}\equiv \ket{1,0},\, \ket{\Phi^{\pm}}\equiv \frac{1}{\sqrt{2}}\(\ket{1,1} \pm \ket{1,-1}\)\,.
\end{align}
Whereas, the state $\ket{\Psi^{-}}$, which can be written as $\ket{S=0,m=0}$ in the collective spin description, does not evolve under Eq.~\eqref{Eq:dotrhoN}.

Therefore, to study the dynamics of a given initial state it is important to decompose it into the states $\ket{S,m}$ (with $S=0$ or $1$) or alternatively Bell states as discussed above. 
The density matrix of such an initial  state can be represented as
\begin{equation}
\rho(t=0) = F \ket{\Psi^{-}}\bra{\Psi^{-}} + {(1-F)}\, \rho_{\rm T}\,.
\label{Eq:DensityInitial}
\end{equation}
Thus, for the initial state~\eqref{Eq:initial}, the  fidelity of the maximally entangled singlet state is  
\begin{equation}
F = \[1 - \sin (2 \theta)\]/2\,.
\label{Eq:fidelity}
\end{equation}  
To recall, $\ket{\Psi^{-}}$ being the singlet state does not evolve under the action of~\eqref{Eq:dotrhoN}, and the triplet state density-matrix $\rho_{\rm T}$ evolves with time; and  in the strong drive limit we obtain~\cite{Agarwal:1977}
\begin{align}
\rho_{\rm T}(\infty)& = \frac{1}{3}\sum_{m=-1,0,1} \ket{1, m}\bra{1,m} \forget{\equiv (1/3)\, {\mathbf I}_3\,,}
\nonumber \\
&\!\!\!\!\!\!\!\!\!\!\!\equiv
\frac{1}{3}\(\ket{\Psi^{+}}\bra{\Psi^{+}} + \ket{\Phi^{+}}\bra{\Phi^{+}} + \ket{\Phi^{-}}\bra{\Phi^{-}}\)\,.
\label{Eq:final}
\end{align}
 
Thus, the collective decay dynamics of a strongly driven two particle system initially in the state $\sin\theta\ket{e_1, g_2}+\cos\theta\ket{g_1, e_2}$ gives the Werner state (see for example~\cite{Bennett:1996}):
\begin{align}
\rho &= F \ket{\Psi^{-}}\bra{\Psi^{-}}+ \frac{(1-F)}{3}\times \nonumber \\
&\qquad\(\ket{\Psi^{+}}\bra{\Psi^{+}} + \ket{\Phi^{+}}\bra{\Phi^{+}} + \ket{\Phi^{-}}\bra{\Phi^{-}}\)\,,
\label{Eq:Werner2}
\end{align}
with $F$ as the pre-chosen probability of the singlet state component. 

To quantify the parameter $\theta$, we note from Eq.~\eqref{Eq:fidelity} that
$\theta =  (1/2) \sin^{-1}(1 - 2 F).$
To obtain purifiable Werner states with fidelity $F>1/2$ \forget{or rather in the range $F \in (1/2,1)$}  it is required that  $\theta \in (\pi/4 + n \pi , 3 \pi/4 + n \pi)$ where $n$ is a non-negative integer.
\forget{The message is that the entanglement fidelity $F>1/2$ of the generated Werner state for $\theta \in (\pi/4,3\pi/4)$ and $\theta \in (5\pi/4,7\pi/4)$ etc. for the initial state $\sin\theta \ket{e_1, g_2} +\cos\theta \ket{g_1, e_2}$ of the two atoms.}

\begin{figure}
\centerline{\includegraphics[scale=0.43]{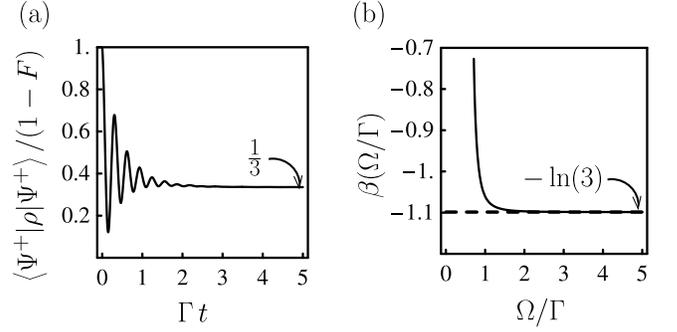}}
\caption{\label{fig:fig}(a) Time evolution of the non-stationary component of the initial state Eq.~\eqref{Eq:DensityInitial}, i.e., $\expect{\Psi^+}{\rho}{\Psi^+}/(1-F)$; it approaches the steady state value $1/3$ for sufficiently strong drive $\Omega/\Gamma =5$ along with the other components $\expect{\Phi^\pm}{\rho}{\Phi^\pm}$ (not plotted here)  as given in Eq.~\eqref{Eq:Werner2}. (b) The non-trivial term in the entropy equation~\eqref{Eq:entropy}---$\beta(\Omega/\Gamma)$---of the steady state. It approaches to that of the Werner state---$\ln(1/3)$---as $\Omega/\Gamma\gg1$ shown by the dashed line.}
\end{figure}

The time evolution of the non-singlet component of the initial state~\eqref{Eq:initial} (a) is shown in Fig.~\ref{fig:fig} to show that it approaches the non-singlet component of the Werner state in Eq.~\eqref{Eq:Werner2}. The von Neumann entropy of the Werner state~\eqref{Eq:Werner2}  is given by
\begin{equation}
S_{\rm Werner}=F \ln F + (1-F)\Bigl[ \ln(1-F) + \ln(1/3)\Bigr]\,.
\label{Eq:EntropyWerner}
\end{equation}
Whereas, the entropy of the general steady state density matrix as a function of a finite parameter $\Omega/\Gamma$ governing the dynamics given by Eq.~\eqref{Eq:dotrhoN} of the initial state Eq.~\eqref{Eq:DensityInitial} can be shown to be
\begin{equation}
S(t=\infty)=F \ln F + (1-F)\Bigl[\ln(1-F) + \beta(\Omega/\Gamma)\Bigr]
\label{Eq:entropy}
\end{equation}
where
$
\beta({\Omega}/{\Gamma}) = \sum_{i} \rho_{i,i} \ln(\rho_{i,i}) \text{ for } i=\{-1,0,1\}.
$
The steady-state density matrix elements are given by~\cite{Agarwal:1977}
\begin{align}
\rho_{1,1} &= \bra{1,1}\rho\ket{1,1}={|\chi|^4}/{D} \nonumber  \\
\rho_{0,0} &= \bra{1,0}\rho\ket{1,0}=-{|\chi|^2(1 - |\chi|^2)}/{D} \nonumber \\
\rho_{-1,-1} &= \bra{1,-1}\rho\ket{1,-1}=1 - \rho_{1,1} - \rho_{0,0}
\end{align}
where $\chi=\ri \sqrt{2}\,\Omega/\Gamma$ and $D= 3 |\chi|^4 - 2 |\chi|^2 +1$.
The function $\beta(\Omega/\Gamma)$ is plotted in Fig.~\ref{fig:fig} (b). It approaches $\ln(1/3)$ so that the entropy~\eqref{Eq:entropy} matches with that of the Werner state given in Eq.~\eqref{Eq:EntropyWerner} as $\Omega/\Gamma\gg1$.

It can be noted that the proposed scheme is very general and can be applied to multipartite systems as well. We start with a general $2N$-partite  density matrix
\begin{equation}
\rho(t=0) = F \ket{\Psi_{\rm s}}\bra{\Psi_{\rm s}} + (1-F) \rho_{\rm T}.
\end{equation}
Here 
$\ket{\Psi_{\rm s}}=\ket{S=0,m=0}$
and for the initial density matrix devoid of the $S=0$ component we assume a very general form given by
\begin{equation}
\rho_T(t=0) = \sum_{S=1}^{N} \sum_{m=-S}^{S}\sum_{m'=-S}^{S} \alpha^{(S)}_{m,m'}\ket{S,m}\bra{S,m'}.
\end{equation}
It needs to be noted at this point that the dynamical equations conserve the spin value $S$. Thus, each spin-multiplet ($m=-S,-S+1,\cdots,S$) evolves on its own without any coupling to other spin counterparts possible for the $2N$ particle state and approaches the state given in Eq.~\eqref{Eq:steadystate} at the steady state in the strong field limit. Thus consolidating all the spin components we obtain the steady state density matrix as
\begin{align}
\rho(t=\infty) &= F \ket{\Psi_{\rm s}}\bra{\Psi_{\rm s}} +(1-F)\times
\nonumber \\
&\qquad\qquad\sum_{S=1}^N\frac{\alpha^{(S)}}{2S +1}\sum_{m=-S}^{S}\ket{S,m}\bra{S,m},
\label{Eq:Nsteadystate}
\end{align}
where \mbox{$\alpha^{(S)} = \sum_{m=-S}^{S} |\alpha_{m,m}^{(S)}|^2$.}
The above state can be recognized as  a $2N$-particle analogue of the Werner state representing an entangled mixed state parametrized by the fidelity of the maximally entangled state. 
Now we use the angular momentum algebra, such that for four particles the individual spin vectors are added in a manner depicted by the notation $\ket{S_1, S_2, [S'], S_3, S_4, [S'']; S, m=m_1+m_2+m_3+m_4}$. Thus, the angular momenta of the first two and last two particles are added together to obtain $S'$ and $S''$ and then combined to obtain total angular momentum quantum numbers $S$ and $m$.  In general, there are several other ways of adding the given number of individual angular momentum vectors, but they are not independent and can be transformed into one another by simple linear transformations. However, consistently using only one such addition scheme is important in one calculation. Using the above addition scheme the various 4-particle spin states can be written out  in terms of the bare atomic states as:
\begin{widetext}
\begin{align}
\forget{\ket{2,2}&\equiv \ket{e_1,e_2,e_3,e_4}\!, \ket{2,-2} \equiv \ket{g_1,g_2,g_3,g_4}\!,
\ket{2,1}\equiv \(\ket{e_1,e_2,e_3,g_4} + \ket{e_1,e_2,g_3,e_4} + \ket{e_1,g_2,e_3,e_4} + \ket{g_1,e_2,e_3,e_4} \)/2\nonumber \\}
\ket{2,0}&\equiv \(\phantom{2}\ket{e_1,e_2,g_3,g_4} + \ket{e_1,g_2,e_3,g_4} +\ket{g_1,e_2,e_3,g_4}+\ket{e_1,g_2,g_3,e_4} +\ket{g_1,e_2,g_3,e_4}
+\phantom{2}\ket{g_1,g_2,e_3,e_4}\)/\sqrt{6} \nonumber 
\\
\ket{0,0}_{1}&\equiv \(2\ket{e_1,e_2,g_3,g_4} -\ket{e_1,g_2,e_3,g_4} -\ket{g_1,e_2,e_3,g_4}-\ket{e_1,g_2,g_3,e_4} -\ket{g_1,e_2,g_3,e_4}
+2\ket{g_1,g_2,e_3,e_4}\)/(2\sqrt{3})\nonumber \\
\ket{0,0}_{2} &\equiv \(\ket{e_1,g_2,e_3,g_4} - \ket{e_1,g_2,g_3,e_4} - \ket{g_1,e_2,e_3,g_4} + \ket{g_1,e_2,g_3,e_4}\)/2 \nonumber \\
\ket{1,0}_{1}&\equiv \(\ket{e_1,e_2,g_3,g_4} - \ket{g_1,g_2,e_3,e_4}\)/\sqrt{2}\nonumber \\ \ket{1,0}_{2,3}&\equiv \(\ket{e_1,g_2,e_3,g_4} \mp \ket{e_1,g_2,g_3,e_4} \pm \ket{g_1,e_2,e_3,g_4} - \ket{g_1,e_2,g_3,e_4}\)/2 \forget{\nonumber \\
\ket{1,0}_{3} &\equiv\( \ket{e_1,g_2,e_3,g_4} + \ket{e_1,g_2,g_3,e_4}-\ket{g_1,e_2,e_3,g_4}-\ket{g_1,e_2,g_3,e_4}\)/2}
\label{Eq:FourParticleAMS}
\end{align}
\end{widetext}
It can be noted that the spin states $\ket{0,0}$ and $\ket{1,0}$ possess multiple representations due to the degeneracies present in the addition of several angular momenta to obtain same resultant. Thus, the state $\ket{S=0,m=0}$ can be obtained by two different angular momentum addition paths such as \mbox{$\ket{0,0}_1=\ket{1/2,1/2,[1],1/2,1/2,[1],0,0}$}, and \mbox{$\ket{0,0}_2=\ket{1/2,1/2,[0],1/2,1/2,[0],0,0}$.} Similarly, \mbox{$\ket{1,0}_1=\ket{1/2,1/2,[1],1/2,1/2,[1],1,0}$,}  \mbox{$\ket{1,0}_2=\ket{1/2,1/2,[1],1/2,1/2,[0],1,0}$}, and \mbox{$\ket{1,0}_3=\ket{1/2,1/2,[0],1/2,1/2,[1],1,0}$.} The set in Eq.~\eqref{Eq:FourParticleAMS}, with trivial additions of the states not shown, is orthogonal and can be used to uniquely represent any given atomic state. 

We consider  a 4-particle initial state, $\ket{\psi_4(0)}=\sin\theta\ket{e_1,e_2,g_3,g_4}+ \cos\theta\ket{g_1,g_2,e_3,e_4}$, and show generation of a generalized Werner state at steady state. This state can be rewritten as
\begin{multline}
\ket{\psi_4(0)}=\sin\theta\[ \frac{1}{\sqrt{6}}\(\sqrt{2}\ket{0,0}_1 + \ket{2,0} + \sqrt{3}\ket{1,0}_1\)\] \\+ \cos\theta\[ \frac{1}{\sqrt{6}}\(\sqrt{2}\ket{0,0}_1 + \ket{2,0} - \sqrt{3}\ket{1,0}_1\)\].
\end{multline}
Thus the initial entanglement fidelity is given by
$F = (1/3)[1 + \sin(2\theta)]$.
which is the same as the maximum entanglement fidelity possible for the generalized Werner state for four particles.
This initial 4-particle state,  at the steady state and in the strong drive field limit, would take the from of Eq.~\eqref{Eq:Nsteadystate} with $N=2$, 
\begin{align}
\alpha^{(1)} &= \frac{3}{2}\frac{[1-\sin(2\theta)]}{[2 -\sin(2\theta)]},\quad
\alpha^{(2)} =\frac{1}{2} \frac{[1 + \sin(2\theta)]}{[2-\sin(2\theta)]},\nonumber \\
\text{and }F&=[1 + \sin(2\theta)]/3
\end{align}
generating a 4-particle Werner state with $\ket{\Psi_{\rm s}}\equiv\ket{0,0}_1$.

In conclusion, we have shown how collective decay dynamics of the system of two level atoms interacting with a leaky cavity develops giving rise to entanglement at steady state. With addition of an extra control, in the form of a strong coherent drive, the dynamics shows further interesting features allowing generation of bipartite mixed entangled states, namely Werner states. The technique can be easily extended to generation of mixed entangled states of a general $2N$ particle system which would be essentially multiparticle analogue of the usual bipartite Werner states. To our knowledge, this is the first proposal showing generation of Werner states with atomic qubits. We further expect to see similar results in other contexts such as collective dephasing in two qubits in quantum-dot systems~\cite{Engel:2005}.

Part of this work was carried out (by K.T.K.) at the Jet Propulsion Laboratory under 
a contract with the National Aeronautics and Space Administration (NASA). K.T.K. acknowledges support from the National Research Council and NASA, Codes Y and S. G.S.A. acknowledges support from National Science Foundation, Grant No. NSF-CCF 0524673.


\end{document}